\documentclass[a4paper,11pt]{article}
\pdfoutput=1 

\usepackage{jcappub} 
\usepackage[T1]{fontenc} 
\usepackage{amsmath}
\usepackage{amsthm}
\usepackage{array}
\usepackage{fancybox}
\usepackage{multirow}
\usepackage{appendix}
\usepackage{epsfig}
\usepackage{xcolor}
\usepackage{subcaption}
\usepackage{graphicx}

\title{\boldmath Fermion dynamics in torsion theories}

\author[a]{J.A.R. Cembranos}
\author[a]{J. Gigante Valcarcel}
\author[b,c]{F.J. Maldonado Torralba}

\affiliation[a]{Departamento de F\'{\i}sica Te\'orica I and IPARCOS, Universidad Complutense de Madrid,\\E-28040 Madrid, Spain.} 
\affiliation[b]{Department of Mathematics and Applied Mathematics, University of Cape Town,\\Rondebosch 7701, Cape Town, South Africa.} 
\affiliation[c]{Van Swinderen Institute, University of Groningen, 9747 AG Groningen, The Netherlands.} 

\emailAdd{cembra@fis.ucm.es}
\emailAdd{jorgegigante@ucm.es}
\emailAdd{fmaldo01@ucm.es}

\abstract{In this work we study the non-geodesical behaviour of particles with spin 1/2 in  Poincar\'e gauge theories of gravity via the WKB method. Within this approach, we calculate the trajectories in a particular Poincar\'e gauge theory, discussing the viability of measuring such a motion.}

\begin{document}

\maketitle

\flushbottom

\section{Introduction}
\label{sec:intro}

There is no doubt that General Relativity (GR) is one of the most successful theories in Physics, with a solid mathematical structure and experimental confirmation~\cite{Wald,Will}. As a matter of fact, we are still measuring for the first time some phenomena that was predicted by the theory a hundred years ago, like gravitational waves~\cite{GW}. Nevertheless it presents some problems that need to be addressed. For example, it cannot be formulated as a renormalizable and unitary Quantum Field Theory. Also, the introduction of spin matter in the energy-momentum tensor of GR may be cumbersome, since we have to add new formalisms, like the spin connection.
These problems can be solved by introducing a gauge approach to the gravitational theories. This task was addressed by Sciama and Kibble in~\cite{Sciama} and~\cite{Kibble}, respectively, where they started to introduce the idea of a Poincar\'e Gauge (PG) formalism for gravitational theories. Following this description one finds that the connection must be compatible with the metric, but not necessarily symmetric. Therefore, it appears a non-vanishing torsion field, that is consequence of the asymmetric character of the connection. For an extensive review of the theories that arise through this reasoning see~\cite{Gauge}. 

Since these kinds of theories were established, there has been a lot of discussion on how would particles behave in a spacetime with a torsion background. In the case of scalar particles, it is clear to see that they should follow geodesics, since the covariant derivative of a scalar field does not depend on the affine connection.
In addition, by assuming the minimum coupling principle, we have that light keeps moving along null geodesics, as in the standard framework of GR. This is because it is impossible to perform the minimally coupling prescription for the Maxwell's field while maintaining the $U\left(1\right)$ gauge invariance~\cite{Hehlspin}. Therefore the Maxwell equations remain in the same form. 
The most differential part occurs when we try to predict how particles with spin $1/2$ should move within this background. This question deserves a deeper analysis, mainly because these kinds of physical trajectories differ from the ones predicted by GR, and if we are able to measure such differences, we will be devising a method to determine the possible existence of a torsion field in our universe. Furthermore, if we know the corresponding equations of motion we can also calculate the strength of this field, although we already have some constraints thanks to torsion pendulums and cosmography observations~\cite{Neil,Alvaro}. In~\cite{ProbeB} we find a comprehensive review of all the proposals that have been made to explain this behaviour. Nevertheless, even nowadays there is no consensus about which one explains it more properly. Here, we will outline the most important suggestions:
\begin{itemize}
\item In 1971, Ponomariev~\cite{Ponomariev} proposed that the test particles move along autoparallels (curves in which the velocity is parallel transported along itself with the total connection). There was no reason given, but surprisingly this has been a recurrent proposal in the posterior literature~\cite{Kleinert, Mao}. 
\item Hehl~\cite{Measure}, also in 1971, obtained the equation of motion via the energy-momentum conservation law, in the single-point approximation, i.e. only using first order terms in the expansion used to solve the energy-momentum equation. He also pointed out that torsion could be measured by using spin $1/2$ particles.
\item In 1981, Audretsch~\cite{WKB} analysed the movement of a Dirac electron in a spacetime with torsion. He employed the WKB approximation, and obtained the same results that Rumpf had obtained two years earlier via an unconventional quantum mechanical approach~\cite{Rumpf}. It was with this article that the coupling between spin and torsion was understood. 
\item In 1991, Nomura, Shirafuji and Hayashi~\cite{Hayashi} computed the equations of motion by the application of the Mathisson-Papapetrou (MP) method to expand the energy-momentum conservation law. They obtained the equations at first order, which are the ones that Hehl had already calculated, but also made the second order approximation, finding the same spin precession as Audretsch.
\end{itemize}
In order to clarify these ideas we organise the article as follows. First, in section~\ref{sec:2} we introduce the mathematical structure of PG theories, and establish the conventions. Then, in the two following sections we review the WKB approximation by Audretsch and the MP approach by Nomura et al., comparing them and presenting the reasons to consider the former for our principal calculations. In the fifth section we present the Raychaudhuri equation in the WKB approximation, and use one of its parameters as an indicator of the strength of the spin-torsion coupling. In section~\ref{sec:cal} we compute the acceleration and the respective trajectories of an electron in a particular solution, and compare it with the geodesical behaviour predicted by GR. The final section is devoted to conclusions and future applications.

\section{Mathematical structure of Poincar\'e gauge theories}
\label{sec:2}

In this section, we give an introduction to the gravitational theories endowed with a non-symmetric connection that still fulfills the metricity condition. The most interesting fact about these theories is that they appear naturally as a gauge theory of the Poincar\'e Group~\cite{Gauge, Shapiro}, making their formalism closer to that of the Standard Model of Particles, therefore postulating it as a suitable candidate to explore the quantization of gravity. We will use the same convention as~\cite{WKB} in order to simplify the discussion.

Since the connection is not necessarily symmetric, the torsion may be different from zero
\begin{equation}
\label{eq:1}
T_{\mu\nu}^{\,\,\,\,\,\,\rho}=\Gamma_{\left[\mu\nu\right]}\,^{\rho}.
\end{equation}
For an arbitrary connection, that meets the metricity condition, there exists a relation with the Levi-Civita connection
\begin{equation}
\label{eq:2}
\mathring{\Gamma}_{\mu\nu}\,^{\rho}=\Gamma_{\mu\nu}\,^{\rho}+K_{\mu\nu}\,^{\rho},
\end{equation}
where 
\begin{equation}
\label{eq:3}
K_{\mu\nu}^{\,\,\,\,\,\,\rho}=T_{\,\,\,\,\nu\mu}^{\rho}+T_{\,\,\,\,\mu\nu}^{\rho}-T_{\mu\nu}^{\,\,\,\,\,\,\rho}
\end{equation}
is the \emph{contortion} tensor. Here, the upper index $\mathring{\,}$ denotes the quantities associated with the Levi-Civita connection.

Since the curvature tensors depend on the connection, there is a relation between the ones defined throughout the Levi-Civita connection and the general ones. For the Riemann tensor we have
\begin{equation}
\label{eq:4}
\mathring{R}_{\mu\nu\rho}^{\,\,\,\,\,\,\,\,\,\,\sigma}=R_{\mu\nu\rho}^{\,\,\,\,\,\,\,\,\,\,\sigma}+\mathring{\nabla}_{\nu}K_{\mu\rho}^{\,\,\,\,\,\,\sigma}-\mathring{\nabla}_{\rho}K_{\mu\nu}^{\,\,\,\,\,\,\sigma}-K_{\alpha\nu}^{\,\,\,\,\,\,\sigma}K_{\mu\rho}^{\,\,\,\,\,\,\alpha}+K_{\alpha\rho}^{\,\,\,\,\,\,\sigma}K_{\mu\nu}^{\,\,\,\,\,\,\alpha}.
\end{equation}
By the usual contractions one can obtain the expressions for the Ricci tensor and Ricci scalar.\\

Here we have just exposed all of these concepts in the usual spacetime coordinates. Nevertheless, it is customary in PG theories to make calculations in the tangent space, that we assume in terms of the Minkowski metric $\eta_{ab}$. At each point of the spacetime we will have a different tangent space, that it is defined through a set of orthonormal tetrads (or \emph{vierbein}) $e_{a}^{\alpha}$, that follow the relations
\begin{equation}
e_{\,\,\,a}^{\mu}e_{\mu b}=\eta_{ab},\,\,\,\,\,\,\,e_{\,\,\,a}^{\mu}e^{\nu a}=g^{\mu\nu},\,\,\,\,\,\,\,e_{\mu}^{\,\,\,a}e_{\nu a}=g_{\mu\nu},\,\,\,\,\,\,\,e_{\mu}^{\,\,\,a}e^{\mu b}=\eta^{ab},
\end{equation}  
where the latin letters refer to the tangent space and the greek ones to the spacetime coordinates. It is clear that if these properties hold, then
\begin{equation}
g_{\mu\nu}=e_{\mu}^{\,\,\,a}e_{\nu}^{\,\,\,b}\eta_{ab}.
\end{equation}
All the calculations from now on will be considered in gravitational theories characterized by this geometrical background.

\section{WKB method}

In this section we summarize the results obtained by Audtresch in~\cite{WKB}, where he calculated the precession of spin and the trajectories of Dirac particles in torsion theories.
The starting point is the Dirac equation of a spinor field minimally coupled to torsion
\begin{equation}
\label{eq:7}
i\hbar\left(\gamma^{\mu}\mathring{\nabla}_{\mu}\Psi+\frac{1}{4}K_{\left[\alpha\beta\delta\right]}\gamma^{\alpha}\gamma^{\beta}\gamma^{\delta}\Psi\right)-m\Psi=0,
\end{equation}
where the $\gamma^{\alpha}$ are the modified gamma matrices, related to the standard ones by the vierbein
\begin{equation}
\gamma^{\alpha}=e^{\alpha}\,_{a}\gamma^{a},
\end{equation}
and $\Psi$ is a general spinor state.

It is worthwhile to note that the contribution of torsion to the Dirac equation is proportional to the antisymmetric part of the torsion tensor, therefore, a torsion field with vanishing antisymmetric component will not couple to the Dirac field. This is usually known as \emph{inert torsion}. 
Since there is no analytical solution to Equation~\eqref{eq:7}, we need to make approximations in order to solve it. As it is usual in Quantum Mechanics, we can use the WKB expansion to obtain simpler versions of this equation.

So, we can expand the general spinor in the following way
\begin{equation}
\label{eq:8}
\Psi\left(x\right)=e^{i\frac{S\left(x\right)}{\hbar}}(-i\hbar)^{n}a_{n}\left(x\right),
\end{equation}
where we have used the Einstein sum convention (with $n$ going from zero to infinity). We have also assumed that $S\left(x\right)$ is real and $a_{n}\left(x\right)$ are spinors.
As every approximation, it has a limited range of validity. In this case, we can use it as long as $\mathring{R}^{-1}\gg \lambda_{B}$, where $\lambda_{B}$ is the de Broglie wavelength of the particle. This constraint expresses the fact that we cannot applied the mentioned approximation in presence of strong gravitational fields and that we cannot consider highly relativistic particles.

If we insert the expansion into the Dirac equation we obtain the following expressions for the zero and first order in $\hbar$:
\begin{equation}
\label{eq:9}
\left(\gamma^{\mu}\mathring{\nabla}_{\mu}S+m\right)a_{0}\left(x\right)=0,
\end{equation}
and
\begin{equation}
\label{eq:10}
\left(\gamma^{\mu}\mathring{\nabla}_{\mu}S+m\right)a_{1}\left(x\right)=-\gamma^{\mu}\mathring{\nabla}_{\mu}a_{0}-\frac{1}{4}K_{\left[\alpha\beta\delta\right]}\gamma^{\alpha}\gamma^{\beta}\gamma^{\delta}a_{0}.
\end{equation}
We then assume that the four-momentum of the particles is orthogonal to the surfaces of constant $S\left(x\right)$, and introduce it as
\begin{equation}
\label{eq:11}
p_{\mu}=-\partial_{\mu}S.
\end{equation}
Then, if we stick to the lowest order, as a consequence of Equation~\eqref{eq:9}, the particles will follow geodesics, as one might expect. But, what happens if we consider the first order in $\hbar$?
For the explicit calculations we refer the reader to~\cite{WKB}, we will just state the definitions and give the main results.

To obtain the equation for spin precession we have considered the spin density tensor as
\begin{equation}
\label{eq:12}
S^{\mu\nu}=\frac{\overline{\Psi}\sigma^{\mu\nu}\Psi}{\overline{\Psi}\Psi},
\end{equation}
where the $\sigma^{\mu\nu}$ are the modified spin matrices, given by
\begin{equation}
\sigma^{\alpha\beta}=\frac{i}{2}\left[\gamma^{\alpha},\,\gamma^{\beta}\right].
\end{equation}
Then, we can obtain the spin vector from this density
\begin{equation}
\label{eq:13}
s^{\mu}=\frac{1}{2}\varepsilon^{\mu\nu\alpha\beta}u_{\nu}S_{\alpha\beta},
\end{equation}
where $\varepsilon^{\mu\nu\alpha\beta}$ is the modified Levi-Civita tensor, related to the usual one by the vierbein
\begin{equation}
\varepsilon^{\mu\nu\alpha\beta}=e_{\,\,\,a}^{\mu}e_{\,\,\,b}^{\nu}e_{\,\,\,c}^{\alpha}e_{\,\,\,d}^{\beta}\varepsilon^{abcd},
\end{equation}
and $u^{\mu}$ represents the velocity of the particle
\begin{equation}
u^{\mu}=\frac{dx^{\mu}}{dt}=x'^{\mu}.
\end{equation}
Via the WKB expansion, we find that we can write the lowest order of the spin vector as
\begin{equation}
\label{eq:14}
s_{0}^{\mu}=\overline{b}_{0}\gamma^{5}\gamma^{\mu}b_{0},
\end{equation}
where $b_{0}$ is the $a_{0}$ spinor but normalised.

With these definitions, we can compute the evolution of the spin vector
\begin{equation}
\label{eq:15}
u^{\alpha}\mathring{\nabla}_{\alpha}s_{0}^{\mu}=3K^{\left[\mu\beta\delta\right]}s_{0\,\delta}u_{\beta}.
\end{equation}
On the other hand, the calculation of the acceleration of the particle comes from the splitting of the Dirac current via the Gordon decomposition and from the identification of the velocity with the normalised convection current. Then it can be shown that the non-geodesical behaviour is governed by the following expression 
\begin{equation}
\label{eq:16}
a_{\mu}=v^{\varepsilon}\mathring{\nabla}_{\varepsilon}v_{\mu}=\frac{\hbar}{4m_{esp}}\widetilde{R}_{\mu\nu\alpha\beta}\overline{b}_{0}\sigma^{\alpha\beta}b_{0}v^{\nu},
\end{equation}
where $\widetilde{R}_{\mu\nu\alpha\beta}$ refers to the intrinsic part of the Riemann tensor associated with the totally antisymmetric component of the torsion tensor:
\begin{equation}
\label{eq:17}
\widetilde{\Gamma}_{\mu\nu}^{\,\,\,\,\,\lambda}=\mathring{\Gamma}_{\mu\nu}^{\,\,\,\,\,\lambda}+3T_{\left[\mu\nu\alpha\right]}g^{\alpha\lambda}.
\end{equation}
Unlike most of the literature exposed in the introduction, the expression \eqref{eq:16} does not have an explicit contortion term coupled to the spin density tensor, hence all the torsion information is encrypted into the mentioned part of the Riemann tensor. Finally, it is worthwhile to note that the standard case of GR is naturally recovered for inert torsion, as expected. \\

\section{Raychaudhuri equation}

One way of studying the consequences of the non-geodesical behaviour is to analyse the evolution of a congruence of the resulting curves throughout the Raychaudhuri equation. Also, this will provide more clues about the singular behaviour of these particles, and will help us to assure previous conclusions reached by the authors in~\cite{JJFJ}.
It is known that Killing vectors define a static frame that will allow us to measure the dynamical quantities with respect to it~\cite{HE}. Nevertheless, in general, an arbitrary spacetime will not have Killing vectors, therefore we do not have a preferred frame to measure the acceleration. In this case, the best one can do is to measure the relative acceleration of two close bodies, which is studied by the analysis of the behaviour of congruences of timelike curves.

In order to observe the evolution of a congruence of curves, we shall study the Raychaudhuri equation in spacetimes with torsion, that has been analysed thoroughly in the literature\footnote{Note that preliminary studies in this subject have not taken into account the change in the deviation vector due to torsion \cite{Kar,Capozziello:2001mq}.} \cite{Luz:2017ldh,Dey:2017fld}.\\
To obtain this equation, we consider the tensor field $B_{\mu\nu}=\tilde{\nabla}_{\nu}v_{\mu}$, which entirely describes the evolution of the separation vector in a congruence of timelike geodesics, calculated in terms of the physical connection $\tilde{\Gamma}$. It is convenient to write this tensor in terms of two components, one orthogonal ($B_{\bot\mu\nu}$), and the other one parallel ($B_{\Vert\mu\nu}$) to the congruence. Given the spatial metric $h_{\mu\nu}$ of the hypersurface orthogonal to the congruence at a given point, one has
\begin{equation}
B_{\mu\nu}=B_{\bot\mu\nu}+B_{\Vert\mu\nu}\,,
\end{equation}
where
\begin{equation}
B_{\bot\mu\nu}=h_{\mu}^{\rho}h_{\nu}^{\sigma}B_{\rho\sigma}\quad {\rm and}\quad B_{\Vert\mu\nu}=B_{\mu\nu}-B_{\bot\mu\nu}\,.
\end{equation}
At the same time, the orthogonal part $B_{\bot\mu\nu}$, can be decomposed into its antisymmetric component $\omega_{\mu\nu}$, known as \emph{vorticity}, a traceless symmetric $\Sigma_{\mu\nu}$, usually referred as \emph{shear}, and its trace $\theta$, also known as \emph{expansion}, such as
\begin{equation}
B_{\bot\mu\nu}=\frac{1}{3}\tilde{\theta} h_{\mu\nu}+\tilde{\Sigma}_{\mu\nu}+\tilde{\omega}_{\mu\nu}\,,
\end{equation}
Then, it can be seen that~\cite{Luz:2017ldh}
\begin{eqnarray}
\label{rayy}
v^{\rho}\tilde{\nabla}_{\rho}\tilde{\theta}&=&\frac{d\tilde{\theta}}{ds}=-\frac{1}{3}\tilde{\theta}^{2}-\tilde{\Sigma}^{\mu\rho}\tilde{\Sigma}_{\mu\rho}-\tilde{\omega}^{\mu\rho}\tilde{\omega}_{\mu\rho}-\tilde{R}_{\rho\varphi}v^{\rho}v^{\varphi}+\tilde{\nabla}_{\mu}\left(v^{\nu}\tilde{\nabla}_{\nu}v^{\mu}\right)
\nonumber
\\
&+&2T_{\beta\alpha}\,^{\rho}v^{\beta}\left(\tilde{\Sigma}_{\ \rho}^{\alpha}+\tilde{\omega}_{\ \rho}^{\alpha}+\frac{1}{3}\tilde{\theta}h_{\ \rho}^{\alpha}+v_{\rho}\tilde{a}^{\alpha}\right)+2v^{\nu}\tilde{\nabla}_{\nu}\left(v^{\mu}T_{\mu\rho}\,^{\rho}\right)\,,
\end{eqnarray}
which is the Raychaudhuri equation. \\

Now, using the fact that the torsion tensor of the connection $\tilde{\Gamma}$ is totally antisymmetric one obtains the following simplification of Equation \eqref{rayy}
\begin{equation}
\label{simp}
\dot{\mathring{\theta}}=\mathring{\nabla}_{\alpha}\mathring{a}^{\alpha}-\mathring{\Sigma}_{\alpha\beta}\mathring{\Sigma}^{\alpha\beta}-\tilde{\omega}_{\alpha\beta}\tilde{\omega}^{\alpha\beta}-\frac{1}{3}\mathring{\theta}^{2}-\mathring{R}_{\alpha\beta}v^{\alpha}v^{\beta}+2T_{\beta\alpha}\,^{\rho}v^{\beta}\tilde{\omega}^{\alpha}\,_{\rho}\,.
\end{equation}
For the next step, we take into account the relation between the vorticity calculated with respect to both connections
\begin{equation}
\tilde{\omega}_{\alpha\beta}=\mathring{\omega}_{\alpha\beta}+2T_{\alpha\beta}\,^{\mu}v_{\mu}\,.
\end{equation}
Then, substituing this relation in \eqref{simp} one obtains
\begin{eqnarray}
\dot{\mathring{\theta}}&=&\mathring{\nabla}_{\alpha}\mathring{a}^{\alpha}-\mathring{\Sigma}_{\alpha\beta}\mathring{\Sigma}^{\alpha\beta}-\left(\mathring{\omega}_{\alpha\beta}+2T_{\alpha\beta}\,^{\mu}v_{\mu}\right)\left(\mathring{\omega}^{\alpha\beta}+2T^{\alpha\beta\lambda}v_{\lambda}\right)-\frac{1}{3}\mathring{\theta}^{2}-\mathring{R}_{\alpha\beta}v^{\alpha}v^{\beta}
\nonumber
\\
&+&2T_{\beta\alpha}\,^{\rho}v^{\beta}\left(\mathring{\omega}^{\alpha}\,_{\rho}+2T^{\alpha}\,_{\rho}\,^{\mu}v_{\mu}\right).
\end{eqnarray}
Then, we use the fact that if we consider a congruence orthogonal to an spacelike hypersurface the Levi-Civita vorticity $\mathring{\omega}$ is null \cite{Poisson:2009pwt}, namely
 \begin{eqnarray}
\dot{\mathring{\theta}}&=&\mathring{\nabla}_{\alpha}\mathring{a}^{\alpha}-\mathring{\Sigma}_{\alpha\beta}\mathring{\Sigma}^{\alpha\beta}-\frac{1}{3}\mathring{\theta}^{2}-\mathring{R}_{\alpha\beta}v^{\alpha}v^{\beta}-4T_{\alpha\beta}\,^{\mu}T^{\alpha\beta\lambda}v_{\mu}v_{\lambda}+4T_{\beta\alpha}\,^{\rho}T^{\alpha}\,_{\rho}\,^{\mu}v^{\beta}v_{\mu}
\nonumber
\\
&=&\mathring{\nabla}_{\alpha}\mathring{a}^{\alpha}-\mathring{\Sigma}_{\alpha\beta}\mathring{\Sigma}^{\alpha\beta}-\frac{1}{3}\mathring{\theta}^{2}-\mathring{R}_{\alpha\beta}v^{\alpha}v^{\beta}\,.
\end{eqnarray}
Moreover, if we substitute the acceleration given in Equation~\eqref{eq:16} into the previous equation, we obtain
\begin{equation}
v^{\rho}\mathring{\nabla}_{\rho}\mathring{\theta}=\frac{d\mathring{\theta}}{ds}=-\frac{1}{3}\mathring{\theta}^{2}-\mathring{\Sigma}^{\mu\rho}\mathring{\Sigma}_{\mu\rho}-\mathring{R}_{\rho\varphi}v^{\rho}v^{\varphi}+\frac{\hbar}{4m_{esp}}\mathring{\nabla}_{\mu}\left(\widetilde{R}_{\,\,\,\nu\alpha\beta}^{\mu}\overline{b}_{0}\sigma^{\alpha\beta}b_{0}v^{\nu}\right).
\end{equation}
Therefore it is clear that in this case the only difference with respect to the geodesical movement is the acceleration term. Let us analyse it in more detail:
\begin{eqnarray}
\label{decomp}
\mathring{\nabla}_{\mu}\left(\widetilde{R}_{\,\,\,\nu\alpha\beta}^{\mu}\overline{b}_{0}\sigma^{\alpha\beta}b_{0}v^{\nu}\right)&=&\left(\mathring{\nabla}_{\mu}\widetilde{R}_{\,\,\,\nu\alpha\beta}^{\mu}\right)\overline{b}_{0}\sigma^{\alpha\beta}b_{0}v^{\nu}+\widetilde{R}_{\,\,\,\nu\alpha\beta}^{\mu}\left[\mathring{\nabla}_{\mu}\left(\overline{b}_{0}\sigma^{\alpha\beta}b_{0}\right)\right]v^{\nu}
\nonumber
\\
&+&\widetilde{R}_{\,\,\,\nu\alpha\beta}^{\mu}\overline{b}_{0}\sigma^{\alpha\beta}b_{0}\mathring{\nabla}_{\mu}v^{\nu},
\end{eqnarray}
where we have used the Leibniz rule for the covariant derivative. Let us study the different contributions separately.

For the third term we have that:
\begin{equation}
\widetilde{R}_{\,\,\,\nu\alpha\beta}^{\mu}\overline{b}_{0}\sigma^{\alpha\beta}b_{0}\mathring{\nabla}_{\mu}v^{\nu}=\widetilde{R}_{\,\,\,\,\,\,\alpha\beta}^{\mu\nu}\overline{b}_{0}\sigma^{\alpha\beta}b_{0}\left(\frac{1}{3}\mathring{\theta} h_{\mu\nu}+\mathring{\Sigma}_{\mu\nu}\right)\,,
\end{equation} 
where the Levi-Civita vorticity tensor is not present in this expression because we are considering a congruence orthogonal to an spacelike hypersurface.
Since the two contracted indexes $\mu$ and $\nu$ of the Riemann tensor are antisymmetric and the tensors $h$ and $\mathring{\Sigma}$ are symmetric we have that:
\begin{equation}
\widetilde{R}_{\,\,\,\nu\alpha\beta}^{\mu}\overline{b}_{0}\sigma^{\alpha\beta}b_{0}\mathring{\nabla}_{\mu}v^{\nu}=0\,.
\end{equation}

In general, for the first and the second term of Expression \eqref{decomp} we cannot find any simplification. In any case, the appearance of focal points will occur when
\begin{equation}
\mathring{R}_{\rho\varphi}v^{\rho}v^{\varphi}\geq A_{\nu}v^{\nu},
\end{equation}
where 
\begin{equation}
A_{\nu}=\frac{\hbar}{4m_{esp}}\mathring{\nabla}_{\mu}\left(\widetilde{R}_{\,\,\,\nu\alpha\beta}^{\mu}\overline{b}_{0}\sigma^{\alpha\beta}b_{0}\right).
\end{equation}
As explained at the beginning of this section, this term gives us the contribution of torsion to the relative acceleration between two spin 1/2 particles, making it a good indicator to  see the difference with respect to a geodesical behaviour. Therefore, we can make a more rigorous approach to the singular behaviour of these particles.
In~\cite{JJFJ} the authors claim that the appearance of n-dimensional black/white hole regions was a good criteria for the occurrence of singularities, even for the Dirac particles, given that the difference with the geodesical movement were not so strong near the event horizon. Now we can say that this will be a good criteria as long as $A_{\nu}\ll 1$, which is what we expect in plausible spacetimes with Dirac particles.

\section{Calculations within the Reissner-Nordstr\"{o}m geometry induced by torsion}
\label{sec:cal}

In this section we will calculate the acceleration and trajectories of electrons in a Reissner-Nordstr\"{o}m solution obtained by two of the authors in the framework of PG field theory of gravity, with the following vacuum action~\cite{JJ,JJ2}:
\begin{eqnarray}
S=\frac{1}{16\pi}\int d^{4}x\sqrt{-g}\left[-\mathring{R}+\frac{d_{1}}{2}R_{\lambda\rho\mu\nu}R^{\mu\nu\lambda\rho}-\frac{d_{1}}{4}R_{\lambda\rho\mu\nu}R^{\lambda\rho\mu\nu} \right.
\nonumber
\\
\left.-\frac{d_{1}}{2}R_{\lambda\rho\mu\nu}R^{\lambda\mu\rho\nu}+d_{1}R_{\mu\nu}\left(R^{\mu\nu}-R^{\nu\mu}\right)\right].
\end{eqnarray}
The exact metric of the solution is
\begin{equation}
ds^{2}=f\left(r\right)dt^{2}-\frac{1}{f\left(r\right)}dr^{2}-r^{2}\left(d\theta^{2}+sin^{2}\theta d\varphi^{2}\right),
\end{equation}
where
\begin{equation}
f\left(r\right)=1-\frac{2m}{r}+\frac{d_{1}\kappa^{2}}{r^{2}}.
\end{equation}
From now on we will consider $d_{1}=1$, which simplifies the computations.\\
In order to know the total and modified connection we need to have the values of the non-vanishing torsion components, which are:
\begin{equation}
\begin{cases}
T_{tr}^{\,\,\,\,\,t}=\frac{a(r)}{2}=\frac{\dot{f}\left(r\right)}{4f\left(r\right)},\\
\,\\
T_{tr}^{\,\,\,\,\,r}=\frac{b(r)}{2}=\frac{\dot{f}\left(r\right)}{4},\\
\,\\
T_{t\theta_{i}}^{\,\,\,\,\,\theta_{i}}=\frac{c(r)}{2}=\frac{f\left(r\right)}{4r},\\
\,\\
T_{r\theta_{i}}^{\,\,\,\,\,\theta_{i}}=\frac{g(r)}{2}=-\frac{1}{4r},\\
\,\\
T_{t\theta_{i}}^{\,\,\,\,\,\theta_{j}}=e^{a\theta_{j}}e_{\,\,\theta_{i}}^{b}\varepsilon_{ab}\frac{d\left(r\right)}{2}=e^{a\theta_{j}}e_{\,\,\theta_{i}}^{b}\varepsilon_{ab}\frac{\kappa}{2r},\\
\,\\
T_{r\theta_{i}}^{\,\,\,\,\,\theta_{j}}=e^{a\theta_{j}}e_{\,\,\theta_{i}}^{b}\varepsilon_{ab}\frac{h\left(r\right)}{2}=-e^{a\theta_{j}}e_{\,\,\theta_{i}}^{b}\varepsilon_{ab}\frac{\kappa}{2rf\left(r\right)},
\end{cases}
\end{equation}
where $i,j=1,2$ with $i\neq j$, and we have made the identification $\left\{ \theta_{1},\,\theta_{2}\right\} =\left\{ \theta,\,\varphi\right\}$. Moreover, $\varepsilon_{ab}$ is the Levi-Civita symbol, and  the dot $\dot{\,}$ means the derivative with respect to the radial coordinate. Also, since the definition of the torsion tensor in the mentioned article differs from our conventions, all the components are divided by 2 with respect to the ones in there.\\
Now, with the components of the metric and the torsion tensors, we can calculate the modified connection and therefore the Riemann tensor of Equation~\eqref{eq:16}, in order to obtain the acceleration. Moreover, we know that the $b_{0}$ and $\overline{b}_{0}$ are the lowest order in $\hbar$ of the general spinor state $\Psi$. Then we can use that the most general form of a positive energy solution of the Dirac equation for $b_{0}$ and $\overline{b}_{0}$ is~\cite{spingr}
\begin{equation}
b_{0}=\left(\begin{array}{c}
cos\left(\frac{\alpha}{2}\right)\\
e^{i\beta}sin\left(\frac{\alpha}{2}\right)\\
0\\
0
\end{array}\right)\,;\,\,\,\,\,\,\,\overline{b}_{0}=\left(\begin{array}{cccc}
cos\left(\frac{\alpha}{2}\right), & e^{-i\beta}sin\left(\frac{\alpha}{2}\right), & 0, & 0\end{array}\right)\,;
\end{equation}
where the angles give the direction of the spin of the particle
\begin{equation}
\overrightarrow{n}=\left(\begin{array}{ccc}
sin\left(\alpha\right)cos\left(\beta\right), & sin\left(\alpha\right)sin\left(\beta\right), & cos\left(\alpha\right)\end{array}\right).
\end{equation}

Before calculating the acceleration, let us use this form of the spinor to calculate the corresponding spin vector. Using Equation~\eqref{eq:14} we have
\begin{equation}
\label{eqn:30}
s^{\mu}=\left(\begin{array}{c}
0\\
\,\\
-sin\left(\alpha\right)cos\left(\beta\right)\sqrt{f\left(r\right)}\\
\,\\
-\frac{sin\left(\alpha\right)sin\left(\beta\right)}{r}\\
\,\\
-\frac{cos\left(\alpha\right)csc\left(\theta\right)}{r}
\end{array}\right)\,;\,\,\,\,s_{\mu}=\left(\begin{array}{cccc}
0, & \frac{sin\left(\alpha\right)cos\left(\beta\right)}{\sqrt{f\left(r\right)}}, & rsin\left(\alpha\right)sin\left(\beta\right), & rsin\left(\theta\right)cos\left(\alpha\right)\end{array}\right).
\end{equation} 

\begin{figure}
\centering
\includegraphics[width=0.65\linewidth]{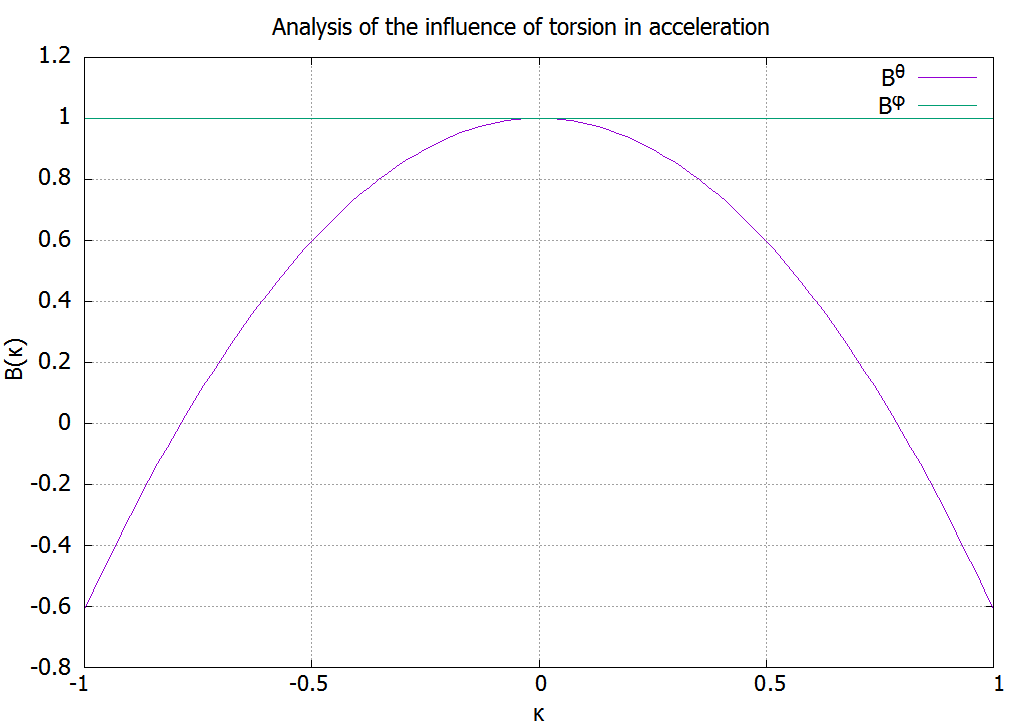}
\label{fig:sfig2}
\caption{We have considered a black hole of 24 solar masses and a particle located near the external event horizon in the $\theta=\pi/2$ plane, at a radial distance of $2m+\varepsilon$, where $\varepsilon=m/10$. The position in $\varphi$ is irrelevant because the acceleration does not depend on this coordinate. For the $B_{\theta}$ case, we assume that the particle has radial velocity equal to 0.8, and that the direction of the spin is in the $\varphi$ direction. The rest of the velocity components are zero except for $v^{t}=(8.8 \kappa +0.3)^{-1/2}$. It is clear from \eqref{plana1} and \eqref{plana2} that we can only calculate the relative acceleration in the $\theta$ direction. For the $B_{\varphi}$ case the velocity is in the $\theta$ direction, and has the same modulus as before. Again, the rest of the components are zero except for 
$v^{t}=1.3 (8.8 \kappa +0.3)^{-1/2}$. The spin has only a radial component, therefore the acceleration would be in the $\varphi$ direction.}
\label{fig:1}
\end{figure}

With all this we can calculate the acceleration for the special case of the solution. To ease the reading of this paper, the acceleration components can be found in the Appendix~\ref{sec:A}.\\
It is worthwhile to note that the only components of the torsion tensor that contribute to the acceleration are those related to the functions $d(r)$ and $h(r)$. This is important, because if we set the $\kappa$ constant to zero, any torsion component does not contribute to the acceleration. Therefore, in this case the torsion tensor is inert, since the axial vector is zero, as expected.\\
On the other hand, The above expressions are complex and  it is difficult to understand their behaviour intuitively. 
In this sense, it is interesting to study two relevant cases that simplify the equations:
\begin{itemize}
\item Low values of $\kappa$:\\
If we consider a realistic physical implementation of this solution, in order to avoid naked singularities, we expect low values of the parameter $\xi=\frac{\kappa}{m^{2}}$. Indeed, $\xi$ is the dimensionless parameter which controls the contribution
of the torsion tensor. Therefore, if we consider the acceleration, we can see that it is a good approximation to consider only up to first order in an expansion of the acceleration in terms of $\xi$. These results can be found in the Appendix~\ref{sec:B}.

\item Asymptotic behaviour:\\
It is interesting to study what happens at the asymptotic limit $r\rightarrow\infty$, in order to observe what is the leading term and compare its strengh with other effects on the particle. We obtain the following:
\begin{eqnarray}
\underset{r\rightarrow\infty}{lim}a^{t}&\simeq&\frac{m^{2}\xi\hbar}{2m_{esp}r}\left(\sin(\alpha)\sin(\beta)\theta'(s)+\sin(\theta)\cos(\alpha)\varphi'(s)\right),
\\
\label{ar}
\underset{r\rightarrow\infty}{lim}a^{r}&\simeq&\frac{m^{2}\xi\hbar}{2m_{esp}r}\left(\sin(\alpha)\sin(\beta)\theta'(s)+\sin(\theta)\cos(\alpha)\varphi'(s)\right),
\\
\underset{r\rightarrow\infty}{lim}a^{\theta}&\simeq&\frac{m\hbar}{2m_{esp}r^{3}}\left[-m\xi r'(s)\left(\sin(\alpha)\sin(\beta)+m^{2}\xi\cos(\alpha)\right)\right.
\nonumber
\\
&+&m\xi t'(s)\left(\sin(\alpha)\sin(\beta)+m^{2}\xi\cos(\alpha)\right)
\nonumber
\\
&-&\left.2\sin(\alpha)\cos(\beta)\sin(\theta)\varphi'(s)\right],
\\
\underset{r\rightarrow\infty}{lim}a^{\varphi}&\simeq&\frac{m\hbar\csc(\theta)}{2m_{esp}r^{3}}\left[m\xi r'(s)\left(m^{2}\xi\sin(\alpha)\sin(\beta)-\cos(\alpha)\right)\right.
\nonumber
\\
&+&\left.m\xi t'(s)\left(\cos(\alpha)-m^{2}\xi\sin(\alpha)\sin(\beta)\right)+2\sin(\alpha)\cos(\beta)\theta'(s)\right].
\end{eqnarray}

Where we have used the viability condition \eqref{via}, because as we will see, that is a neccesary condition for the semiclassical aproximation. \\
We can observe that the time and radial components follow a $r^{-1}$ pattern, while the angular components follow a $r^{-3}$ behaviour. Hence, in the first components the torsion effect goes asymptotically to zero at a lower rate than the strength provided by the conventional gravitational field. Meanwhile in the angular ones, it goes at a higher rate.
\end{itemize} 
It is interesting to analyse the two components of the acceleration that are non-zero in GR, $a^{\theta}$ and $a^{\varphi}$, to reach a deeper understanding. They read
\begin{equation}
\label{plana1}
a^{\theta}|_{\kappa=0}=\frac{m\hbar\sin(\theta)}{2m_{esp}r^{3}\sqrt{1-\frac{2m}{r}}}\left(s^{\varphi}r'(s)+2s^{r}\varphi'(s)\right),
\end{equation}
and
\begin{equation}
\label{plana2}
a^{\varphi}|_{\kappa=0}=\frac{m\hbar\csc(\theta)}{2m_{esp}r^{3}\sqrt{1-\frac{2m}{r}}}\left(s^{\theta}r'(s)+2s^{r}\theta'(s)\right),
\end{equation}
where we have used the expression of the spin vector~\eqref{eqn:30} to simplify the equations. As we can see, the form of the two equations is very similar, and can be made equal by establishing the identifications $\sin(\theta)\leftrightarrow \csc(\theta)$, and $\varphi\leftrightarrow \theta$. 
For two of them we observe that the spin-gravity coupling acts as a {\it cross-product force}, in the sense that the acceleration is perpendicular to the direction of the velocity and the spin vector. \\
Now, to measure the torsion contribution in the acceleration we shall compare the acceleration for $\kappa=0$ and for arbitrary values of $\kappa$. In this sense, we define a new dimensionless parameter as the fraction between the acceleration for a finite value of $\kappa$ and the one given by $\kappa =0$:
\begin{equation}
B^{\mu}(\kappa)=\frac{a^{\mu}}{a^{\mu}|_{\kappa=0}}.
\end{equation}
As we have stated before, the viability condition \eqref{via} implies that 
\begin{equation}
\cos(\alpha)\theta'(s)-\sin(\alpha)\sin(\beta)\sin(\theta)\varphi'(s)=0, 
\end{equation}
so $a^{t}|_{\kappa=0}$ and $=a^{r}|_{\kappa=0}$ vanish identically. This means that we cannot study these two components of the $B^{\mu}$ parameter. Nevertheless, we can still measure it in the angular coordinates.\\
Let us explore two examples, that are shown in Figure~\ref{fig:1}. There we represent different components of 
$B^{\mu}$ in function of $\kappa$ for a fixed position and two different spin and velocity directions.

As can be seen, this gives rise to some interesting features, that we would like to address. First of all, it is worthwhile to stress that there is nothing in the form of the metric or in the underlying theory that stops us from taking negative values of $\kappa$, in contrast with the usual electromagnetic version of the solution. 
We can observe that as we take higher absolute values for $\kappa$ we find that the acceleration caused by the spacetime torsion is directed in the opposite direction of the one produced by the gravitational coupling, reaching significant differences for large $\kappa$. This is expected since we have chosen a strong coupling between spin and torsion. \\

\begin{figure}
\begin{subfigure}{.5\linewidth}
\centering
\includegraphics[width=1\linewidth]{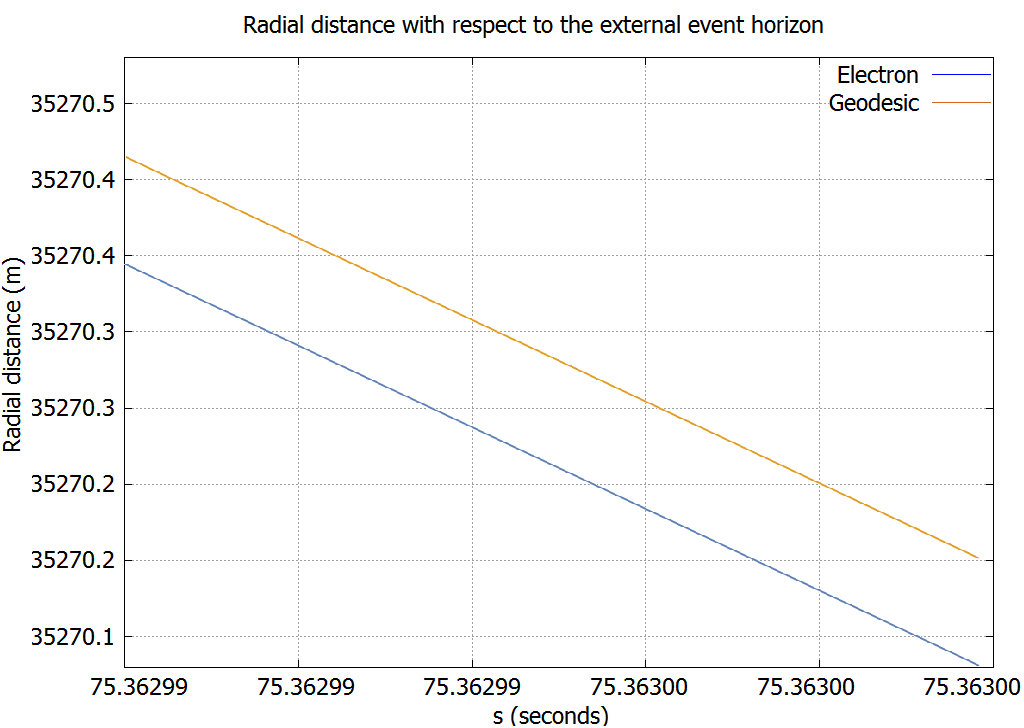}
\caption{Trajectory at 35 km of the event horizon.}
\end{subfigure}%
\begin{subfigure}{.5\linewidth}
\centering
\includegraphics[width=1\linewidth]{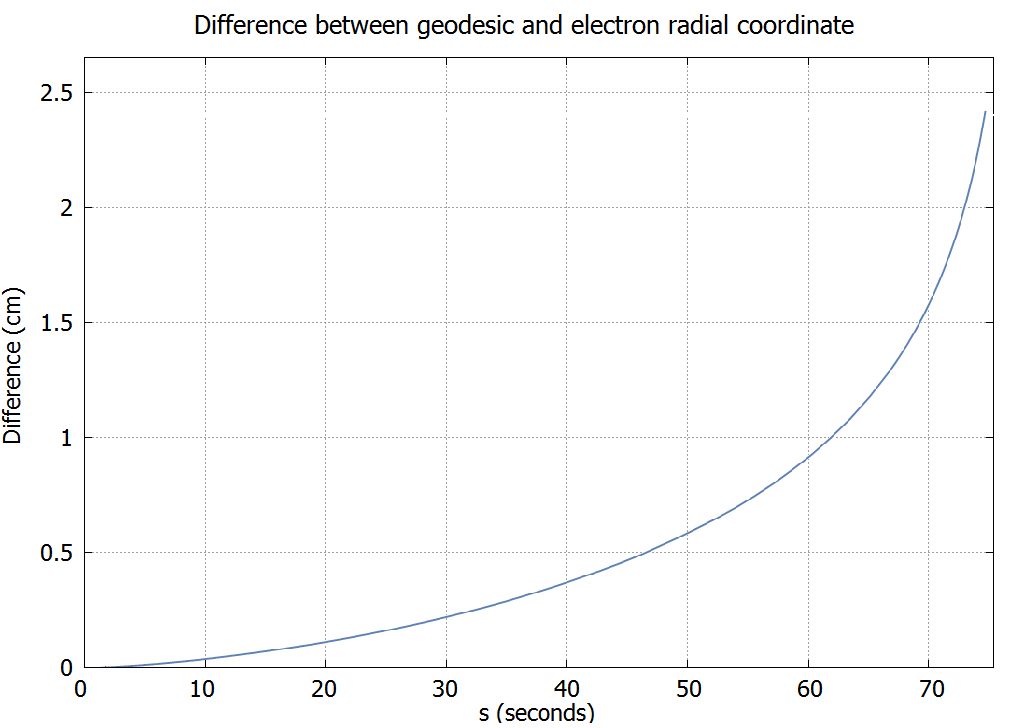}
\caption{Relative position between the two particles.}
\end{subfigure}
\caption{For this numerical computation we have used a black hole with 24 solar masses and $\kappa=10$, with the electron located outside the external event horizon in the $\theta=\pi/2$ plane. We have assumed an electron with radial velocity of 0.9 and initial spin aligned in the $\varphi$ direction. All the rest of the initial conditions are the same than the ones
presented in Figure~\ref{fig:1}.}
\label{fig:3}
\end{figure}

Now, we go one step forward and calculate the trajectory of the particle, using Equation~\eqref{eq:16} and having in mind the spinor evolution equation~\eqref{eq:15}, which can be rewritten as
\begin{equation}
v^{\mu}\tilde{\nabla}_{\mu}b_{0}=0.
\end{equation}
For the exact Reissner-Nordstr\"{o}m geometry supported by torsion, we find several interesting features.
First, in order to maintain the semiclassical approximation and the positive energy associated with the spinor, 
two conditions must be fulfilled:
\begin{equation}
\dot{f} \left(r\right)\ll Lf\left(r\right),
\end{equation}
where $L=3.3\cdot 10^{-8}\,\,m^{-1}$, so that in the units we are using the derivative of $f\left(r\right)$ is at least two orders of magnitude below the value of $f\left(r\right)$.\\
The other one is
\begin{equation}
\label{via}
\left(\overline{b}_{0}\sigma^{r\beta}b_{0}\right)v_{\beta}=0.
\end{equation}
The first one is a consequence of the method that we are applying: if both curvature and torsion are strong then the interaction is also strong, and the WKB approximation fails. This one is a purely metric condition, since it comes from the Levi-Civita part of the Riemann tensor, so it will be the same for all the spherically symmetric solutions. The second one is the radial component of the Pirani condition, that was explained in section 4.
We have solved the above equations numerically for different scenarios, obtaining the results that are shown in Figure~\ref{fig:3}.\\

We have chosen the same trajectories analysed in the discussion of the acceleration. That discussion shows that any difference from the geodesical behaviour in the radial coordinate would be an exclusive consequence of the torsion-spin coupling, with no presence of GR terms, since the acceleration term in this coordinate depends on $\kappa$. Indeed it is possible to have situations under which the geodesics and the trajectories of spin 1/2 particles are distanced due to this effect, even by starting at the same point. If we are able to measure such a difference experimentally, we could have an idea of the specific values of the torsion field present in this particular geometry.

\section{Conclusions}

Motivated by the lack of consensus on how Dirac particles propagate in torsion theories, we review the two main formulations for this purpose and compare them. We reach the conclusion that the WKB method is more consistent for the mentioned task, since it does not need any additional condition, like the Pirani one, in order to solve the resulting equations. In addition,  it seems a better approach to treat the intrinsic spin dynamic from the Dirac equation than from a classical equation like the MP one.\\

After that, we have written the Raychaudhuri equation for the spin particles and defined a new parameter to measure the non-geodesical behaviour. In contrast with just the acceleration given by Equation~\eqref{eq:16},
this parameter constitutes a well-defined physical criterion in order to distinguish observationally the existence
of a non-zero torsion, since it quantifies the difference of the acceleration with respect to the geodesical one 
measured by nearby observers. \\

Finally, we have applied the WKB method to a specific geometrical solution of PG gravity and analysed the results. Within the asymptotic behaviour at large distances, where the WKB approximation holds, the torsion effects are typically much smaller than the contribution given by the Levi-Civita connection. Therefore, it is interesting to find scenarios where this component is not present. In this particular case, we have found a {\it cross-product behaviour} of the gravitational interaction, i.e. an acceleration induced that is perpendicular to the spin direction of the particle and to its velocity when torsion is absent. Therefore differences from geodesical behaviours in other directions can only be consequence of the torsion contribution.

With this fact in mind, we have found a situation where we can appreciate qualitative differences between the geodesical movement and the trajectories of spin 1/2 particles, as shown in Figure~\ref{fig:3}. However, this different dynamics 
needs an important magnitude of the torsion coupling in order to be observed. To have a realistic situation that can be explained through the studied metric, we would need a neutron-star like system, where we have a large concentration of spin aligned particles due to a magnetic field inside the star. In such a case, we could try to observe the difference of angles between photons and neutrinos coming from the same source behind the neutron star. This and other studies will be analysed in future works following the computations developed in this article.

\appendix
\section{Acceleration components}
\label{sec:A}

Here we present the components of the acceleration calculated following the prescription discussed in section~\ref{sec:cal}.

\begin{eqnarray}
a^{t}&=&-\frac{\kappa\hbar}{2m_{esp}r^{2}\left(\frac{\kappa-2mr+r^{2}}{r^{2}}\right)^{3/2}}\left\{ \sqrt{\frac{\kappa-2mr+r^{2}}{r^{2}}}\sin(\alpha)\cos(\beta)r'(s) \right.
\nonumber
\\
&-&\theta'(s)\left[\sin(\alpha)\sin(\beta)\left(r-m\right)+\kappa r\cos(\alpha)\right]
\nonumber
\\
&+&\Biggl.\sin(\theta)\varphi'(s)\left[\cos(\alpha)\left(m-r\right)+\kappa r\sin(\alpha)\sin(\beta)\right]\Biggr\}
\end{eqnarray}

\begin{eqnarray}
a^{r}&=&-\frac{\hbar}{2m_{esp}r^{4}\left(\kappa-2mr+r^{2}\right)}\left\{ r\sqrt{\frac{\kappa-2mr+r^{2}}{r^{2}}}\left[\theta'(s)\left(\cos(\alpha)\left(2m^{2}r^{2}-mr^{3}-3m\kappa r+\kappa^{2}\right.\right.\right.\right.
\nonumber
\\
&-&\left.\left.\kappa^{2}r^{4}+\kappa r^{2}\right)+\kappa r^{3}\sin(\alpha)\sin(\beta)(m-r)\right)+\sin(\theta)\varphi'(s)\left(\sin(\alpha)\sin(\beta)\left(-2m^{2}r^{2}+mr^{3}\right.\right.
\nonumber
\\
&+&\left.\left.\left.3m\kappa r-\kappa^{2}+\kappa^{2}r^{4}-\kappa r^{2}\right)+\kappa r^{3}\cos(\alpha)(m-r)\right)\right]
\nonumber
\\
&+&\Biggl.\kappa\sin(\alpha)\cos(\beta)\left(\kappa-2mr+r^{2}\right)^{2}t'(s)\Biggr\},
\end{eqnarray}

\begin{eqnarray}
a^{\theta}&=&-\frac{\hbar\sin(\theta)}{4m_{esp}r^{7}\left(\frac{\kappa-2mr+r^{2}}{r^{2}}\right)^{3/2}}\Biggl\{ -2\csc(\theta)r'(s)\left[\cos(\alpha)\left(2m^{2}r^{2}-mr^{3}-3m\kappa r+\kappa^{2}-\kappa^{2}r^{4}+\kappa r^{2}\right)\right.\Biggr.
\nonumber
\\
&+&\left.\kappa r^{3}\sin(\alpha)\sin(\beta)(m-r)\right]-2r\left(-\kappa+2mr-r^{2}\right)\left[\sin(\alpha)\cos(\beta)(2mr-\kappa)\sqrt{\frac{\kappa-2mr+r^{2}}{r^{2}}}\varphi'(s)\right.
\nonumber
\\
&-&\Biggl.\Biggl.\kappa\csc(\theta)t'(s)\left(\sin(\alpha)\sin(\beta)(r-m)+\kappa r\cos(\alpha)\right)\Biggr]\Biggr\},
\end{eqnarray}

\begin{eqnarray}
a^{\varphi}&=&-\frac{\hbar\csc(\theta)}{4m_{esp}r^{7}\left(\frac{\kappa-2mr+r^{2}}{r^{2}}\right)^{3/2}}\Biggl\{ 2r'(s)\left[\sin(\alpha)\sin(\beta)\left(2m^{2}r^{2}-mr^{3}-3m\kappa r+\kappa^{2}-\kappa^{2}r^{4}+\kappa r^{2}\right)\right.\Biggr.
\nonumber
\\
&-&\left.\kappa r^{3}\cos(\alpha)(m-r)\right]+2r\left(\kappa-2mr+r^{2}\right)\left[\sin(\alpha)\cos(\beta)(\kappa-2mr)\sqrt{\frac{\kappa-2mr+r^{2}}{r^{2}}}\theta'(s)\right.
\nonumber
\\
&+&\Biggl.\Biggl.\kappa t'(s)\left(\cos(\alpha)(m-r)+\kappa r\sin(\alpha)\sin(\beta)\right)\Biggr]\Biggr\}
\end{eqnarray}

\section{Acceleration at low $\kappa$}
\label{sec:B}

Here we display the acceleration components at first order of the dimensionless parameter $\xi=\kappa/m^{2}$, as indicated in section~\ref{sec:cal}.

\begin{eqnarray}
a^{t}=-\frac{\xi m^{2}\hbar}{2\left(m_{esp}r(r-2m)\sqrt{1-\frac{2m}{r}}\right)}\left[\sin(\alpha)\cos(\beta)\sqrt{1-\frac{2m}{r}}r'(s)\right.
\nonumber
\\
+\Biggl.\left(m-r\right)\left(\sin(\alpha)\sin(\beta)\theta'(s)+\cos(\alpha)\sin(\theta)\varphi'(s)\right)\Biggr]+O\left(\xi^{2}\right),
\end{eqnarray}

\begin{eqnarray}
a^{r}&=&\frac{m\hbar\sqrt{1-\frac{2m}{r}}}{2m_{esp}r^{2}}\left(\cos(\alpha)\theta'(s)-\sin(\alpha)\sin(\beta)\sin(\theta)\varphi'(s)\right)
\nonumber
\\
&-&\frac{\xi m^{2}\hbar}{4\left(m_{esp}r^{4}\sqrt{1-\frac{2m}{r}}\right)}\Biggl[\theta'(s)\left(2r^{2}\sin(\alpha)\sin(\beta)(m-r)+\cos(\alpha)(2r-5m)\right)\Biggr.
\nonumber
\\
&+&\sin(\theta)\varphi'(s)\left(2r^{2}\cos(\alpha)(m-r)+\sin(\alpha)\sin(\beta)(5m-2r)\right)
\nonumber
\\
&+&\left.2r\sin(\alpha)\cos(\beta)\sqrt{1-\frac{2m}{r}}(r-2m)t'(s)\right]+O\left(\xi^{2}\right),
\end{eqnarray}

\begin{eqnarray}
a^{\theta}&=&-\frac{m\hbar}{2m_{esp}r^{4}}\left(\frac{\cos(\alpha)r'(s)}{\sqrt{1-\frac{2m}{r}}}+2r\sin(\alpha)\cos(\beta)\sin(\theta)\varphi'(s)\right)
\nonumber
\\
&+&\frac{m^{2}\hbar\xi}{4m_{esp}r^{5}(r-2m)\sqrt{1-\frac{2m}{r}}}\Biggl[r'(s)\left(2r^{2}\sin(\alpha)\sin(\beta)(m-r)+\cos(\alpha)(2r-3m)\right)\Biggr.
\nonumber
\\
&+&\left.r\sin(\alpha)(r-2m)\left(2\cos(\beta)\sin(\theta)\sqrt{1-\frac{2m}{r}}\varphi'(s)-2\sin(\beta)(m-r)t'(s)\right)\right]
\nonumber
\\
&+&O\left(\xi^{2}\right),
\end{eqnarray}

\begin{eqnarray}
a^{\varphi}&=&\frac{m\hbar\sin(\alpha)\csc(\theta)}{2m_{esp}r^{4}}\left(\frac{\sin(\beta)r'(s)}{\sqrt{1-\frac{2m}{r}}}+2r\cos(\beta)\theta'(s)\right)
\nonumber
\\
&+&\frac{m^{2}\hbar\xi\csc(\theta)}{4m_{esp}r^{5}\sqrt{1-\frac{2m}{r}}(r-2m)}\Biggl[r'(s)\left(2r^{2}\cos(\alpha)(m-r)+\sin(\alpha)\sin(\beta)(3m-2r)\right)\Biggr.
\nonumber
\\
&+&\left.r(r-2m)\left(-2\sin(\alpha)\cos(\beta)\sqrt{1-\frac{2m}{r}}\theta'(s)-2\cos(\alpha)(m-r)t'(s)\right)\right]
\nonumber
\\
&+&O\left(\xi^{2}\right).
\end{eqnarray}

\acknowledgments

This work was partly supported by the projects FIS2014-52837-P (Spanish MINECO), FIS2016-78859-P (AEI/FEDER, UE), Spanish Red Consolider MultiDark FPA2017-90566-REDC. FJMT acknowledges financial support from the National Research Foundation grants 99077 2016-2018 (Ref. No. CSUR150628121624), 110966 (Ref. No. BS170509230233), and the NRF IPRR (Ref. No. IFR170131220846). FJMT acknowledges nancial support from the Erasmus+ KA107 Alliance4Universities programme and from the Van Swinderen Institute at the University of Groningen.


\end{document}